\documentstyle[editedvolume,numreferences]{crckapb} 

\input epsfig

\begin{opening}

\title{How to Count the States of Extremal \protect \\
       Black Holes in $N=8$ Supergravity}

\subtitle{Contribution to the proceedings of the 1997 Cargese summer
school. \\ Preprint HUTP-97/A098, hep-th/9712215}

\author{Vijay Balasubramanian}

\institute{Harvard University, \\
           Lyman Laboratory of Physics, \\
           Cambridge,  MA 02138, \\
           U.S.A.\\
           email: vijayb@pauli.harvard.edu}
\end{opening}

\runningtitle{Counting States of Black Holes}

\newcommand{\vol}{{\rm Vol}}
\newcommand{\tf}[2]{dz^{#1}\wedge d\bar{z}^{\bar{#2}}}

\newcommand{\J}{k}

\begin{document}
\begin{abstract}
$N=8$ supergravity has a rich spectrum of black holes charged under the 56
$U(1)$ gauge fields of the theory.  Duality predicts that the entropy of
these black holes is related to the quartic invariant of the $E(7,7)$
group. We verify this prediction in detail by constructing black holes that
correspond to supersymmetric bound states of 2-branes at angles and
6-branes.  The general bound state contains an arbitrary number of branes
rotated relative to each other, and we derive the condition for these
rotations to preserve supersymmetry.  The microscopic bound state
degeneracy matches the black hole entropy in detail.  The entire 56 charge
spectrum of extremal black holes in $N=8$ supergravity can be displayed as
the orbit under duality of a five parameter generating solution.  We exhibit
a new generating configuration consisting of D3-branes at angles
and discuss its entropy.
\end{abstract}

\section{Introduction}\index{Black hole entropy}
\index{Black holes!four dimensional} \index{Black holes!entropy}
\index{Black holes!$N=8$ supergravity} The basic technique for counting the
states of extremal black holes in string theory is to represent the black
hole as a bound state of p-brane solitons.  The horizon area of the
resulting supergravity solution is related to an entropy by the
Bekenstein-Hawking formula $S = A/4G_N$.  At weak coupling, p-branes
charged under the Ramond-Ramond fields of Type II string theory are easily
quantized as D-branes~\cite{polch95a}.  This fact was exploited in the
seminal paper of Strominger and Vafa who showed that for some black holes
corresponding at weak coupling to systems of D-branes, the collective
coordinate degeneracy of the bound state can be easily
calculated~\cite{sv1}.  If the bound state in question is a BPS state, this
degeneracy can be extrapolated back to strong coupling since the
representation theory of supersymmetry protects it from corrections.  To
summarize, the program for counting the states of black holes amounts to
treating the black hole as a bound state of solitons, quantizing the
solitons at weak coupling, and extrapolating the resulting collective
coordinate degeneracy back to strong coupling via an appeal to
supersymmetry.

\begin{figure}[h]
\begin{center}
\epsfig{figure=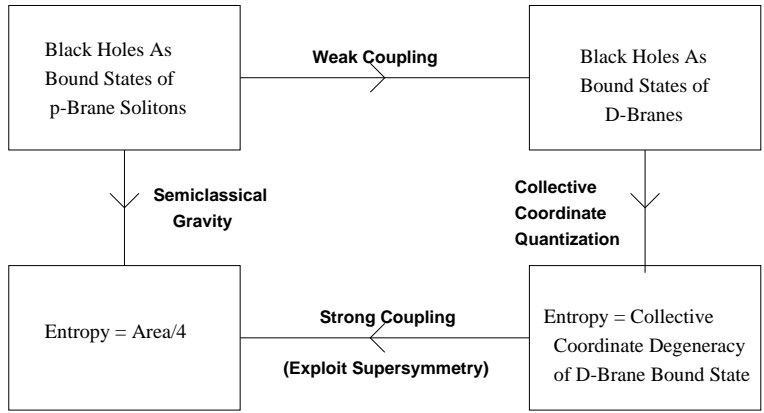, width=0.8\textwidth}
\end{center}
\caption{How to count black hole states\label{fig:statecount}}
\end{figure}

In the present paper we apply this method of counting states (see
Figure~\ref{fig:statecount}) to supersymmetric black holes in $N=8$
supergravity, the low energy effective theory of Type II string theory
compactified on a 6-torus.  In the bulk of the paper we will work with
IIA string theory which has 0-, 2-, 4- and 6-brane solitons.  Charged
black holes can be constructed by wrapping these branes on the
6-torus.  We are interested in supersymmetric black holes because the
weak-to-strong coupling transformation used in
Fig.~\ref{fig:statecount} generally preserves only the degeneracy of
states that are annihilated by some of the supercharges.  Such
supersymmetric black holes are also extremal because they will satisfy
a BPS mass formula relating the mass of the black hole to the charges
it carries.

We begin in Sec.~\ref{sec:n8} by discussing the general prediction for the
entropy of extremal black holes in terms of the quartic invariant of the
$E(7,7)$ duality group of $N=8$ supergravity.  We want to construct
extremal black holes out of bound states of branes.  So, in
Sec.~\ref{sec:angles}, we show how to create general supersymmetric bound
states of 2-branes at angles and 6-branes wrapped on the internal 6-torus.
These are T-dual to bound states of 4-branes at angles and 0-branes.
Sec.~\ref{sec:classical} finds the classical black hole solutions to $N=8$
supergravity corresponding to such branes at angles.  Their classical
entropy $S = A/4G_N$ precisely matches the prediction given by the $E_7$
quartic invariant in Sec.~\ref{sec:n8}.  Then, in Sec.~\ref{sec:counting},
we count the degeneracy of the microscopic bound state of D-branes and
demonstrate detailed agreement with the classical entropy.

In general, extremal black holes in $N=8$ supergravity can carry 56
different $U(1)$ charges.  As discussed in Sec.~\ref{sec:n8}, this
space of black holes can be generated by dualities from a 5-parameter
generating solution~\cite{hull96}.  The known generating solution of
Cvetic and Tseytlin~\cite{ct2} contains NS 5-branes and fundamental
strings and it is difficult to count its states microscopically.  On
the other hand, all treatments in the literature of four dimensional
black hole entropy using D-branes correspond to 4 parameter generating
solutions (for example, see~\cite{strom96d,vf}).  In the
discussion concluding this paper, we show how a 5 parameter generating
solution containing only D-branes can be constructed by intersecting
3-branes at angles.  This solution can be dualized into a system of
4-branes, 2-branes and 0-branes, and clearly displays in D-brane
language the final parameter missing from the discussion of extremal
black hole entropy in $N=8$ supergravity.  Counting the states of
these black holes introduces some interesting new features.

\section{Black Hole Entropy in $N=8$ Supergravity}
\label{sec:n8}
\index{E(7,7)}
The no hair theorem for black holes says that a non-rotating black
hole is completely characterized by its mass and charges.  The mass of
a supersymmetric black hole is further related to its charges by a BPS
bound.  It follows from this that the area and entropy of an extremal
black hole in $N=8$ supergravity should be completely specified by its
charges under the 56 $U(1)$ fields of the theory.  $N=8$ supergravity
has a duality group - $E(7,7)$ - that mixes up these charges and
dresses them with the moduli associated with the shape of the internal
$T^6$ at infinity.  However, the entropy, $S=A/4G_N$, should be
invariant under duality since we do not expect the degeneracy of
supersymmetric states to change in a dual
description~\cite{structure,kallosh96b,kallosh96a}.  So the entropy of
a supersymmetric black hole is expected to be a duality invariant.

The charges of $N=8$ supergravity rotate in the 56 dimensional
antisymmetric tensor representation of the $E(7,7)$ duality
symmetry.\footnote{The useful reference for the remainder of this section
is Cremmer and Julia~\cite{cremmer}.}  These charges are associated with
various solitons wrapped on the internal $T^6$ - for example, there are 15
charges coming from 2-branes wrapped on different cycles and 15 from
4-branes.  The integral quantized charges are generally ``dressed'' by
moduli scalar fields that represent the shape of the asymptotic internal
torus and therefore parametrize inequivalent vacua.  The moduli parametrize
the coset $E(7,7)/SU(8)$ and therefore $SU(8)$ is the part of the duality
group that mixes the charges in a nontrivial way.

For our purposes it is convenient to work in the $SO(8)$ formalism
where the 56 charges are assembled in an 8 by 8 antisymmetric tensor
whose indices rotate in the vector representation of $SO(8) \in
SU(8)$.  Then the charge matrix is:
\begin{equation}
z_{ij} = {1 \over \sqrt{2}} (x_{ij} + i y_{ij})
\end{equation}
For solutions that contain only Ramond-Ramond charges associated with
D-branes, the $x$ and $y$ variables are directly related to wrapped
branes.  Taking $q_{ij}$ and $p_{ijkl}$ to be the charges of 2-branes
and 4-branes wrapped on the $(ij)$ 2-cycle and the $(ijkl)$ 4-cycle,
$x_{ij} = \epsilon^{ijklmn} p_{klmn}/\sqrt{2}$ and $y_{ij} =
q_{ij}/\sqrt{2}$ for $i,j \leq 6$.  Finally, if $Q_0$ and $Q_6$ are
the 0-brane and 6-brane charges in the system, $x^{78} = Q_0/\sqrt{2}$
and $y^{78} = -Q_6/\sqrt{2}$.

By the above arguments, the area of a black hole must be given in
terms of a duality invariant constructed from the charges.  After
accounting for dimensions, this was identified in~\cite{kallosh96a} as
$A = 4\pi\sqrt{J_4}$ where $J_4$ is the quartic
invariant~\cite{cremmer}:
\begin{equation}
-J_4 = x^{ij} y_{jk} x^{kl} y_{li} - 
 {x^{ij} y_{ij} x^{kl} y_{kl} \over 4} +
{\epsilon_{ijklmnop} \over 96}
 (x^{ij} x^{kl} x^{mn} x^{op} +
                                  y^{ij} y^{kl} y^{mn} y^{op})
\label{eq:pred}
\end{equation}
This gives a prediction from duality for the area of four dimensional
black holes in string theory.   In this paper we will construct rather
complicated bound states made from arbitrary numbers of 2-branes and
6-branes and show that the above prediction is verified in detail.

\subsection{Generating Solutions}
\label{sec:gen}
\index{Black holes!duality}
Since a supersymmetric black hole is completely specified by its
charges, and duality mixes up these charges, the entire 56 dimensional
spectrum of black holes can be generated by duality from a much
smaller space of generating solutions.  In fact, the generating
solution must have five independent parameters~\cite{hull96}. To see
this it is convenient to rewrite the charge matrix in $SU(8)$
formalism~\cite{cremmer}:
\begin{equation}
{\cal Z}_{AB} = -{1\over 4} z_{ij} (\Gamma^{ij})_{AB}
\label{eq:SU8form}
\end{equation}
where $\Gamma^{ij}$ are generators of $SO(8)$.  Now ${\cal Z}$
transforms under $SU(8)$ as ${\cal Z} \rightarrow U {\cal Z} U^T$.
The $SU(8)$ transformations can be used to skew diagonalize ${\cal Z}$
giving four complex eigenvalues:
\begin{equation}
{\cal Z} =  \pmatrix{\lambda_1 e^{i\theta_1} \tau & 0 & 0 & 0 \cr
                     0 & \lambda_2 e^{i\theta_2} \tau & 0 & 0 \cr
                     0 & 0 & \lambda_3 e^{i\theta_3} \tau & 0 \cr
                     0 & 0 & 0 & \lambda_4 e^{i\theta_4} \tau}
~~~~~~;~~~~~~\tau=\pmatrix{0 & 1 \cr -1 & 0}
\end{equation}
We can perform further $SU(8)$ transformations that eliminate the phase of
the eigenvalues in first three blocks, by adding compensating phases in the
last block.  So, in general, we can diagonalize to get 1 complex and 3 real
skew eigenvalues for a total of 5 parameters.  Turning this around, in
order to obtain the full spectrum of black holes by duality transformations
of a generating solution, the latter must have five parameters.

The invariant phase that the $SU(8)$ transformations are unable to get
rid of is associated with the presence of electric and magnetic dual
objects at the same time.\footnote{For example, consider a system with
6-branes, and 2-branes on the $(12)$, $(34)$ and $(56)$ cycles.  Then
if we could add 4-brane charge on the $(3456)$ cycle, $z_{12}$ would
be complex.  The resulting phase in ${\cal Z}$ could not be removed by
duality.}  It is difficult to turn on such dual charges
supersymmetrically and so the only five parameter generating solution
available at present is that of~\cite{ct2}.  The bulk of this paper
deals with solutions carring pure electric or magnetic charge; under
duality these transform into a 55 parameter family of solutions, one
short of the desired total.  The conclusion discusses a new five
parameter generating configuration made purely from D3-branes at
angles.

\section{Supersymmetry and Branes at Angles}
\label{sec:angles}
\index{Branes at angles}
\index{Supersymmetry!intersecting branes}
In the previous section we discussed the prediction from duality for
the area of extremal black holes in $N=8$ supergravity:
$A=4\pi\sqrt{J_4}$.  To test this prediction we will construct
complicated black holes containing many microscopic branes and compute
their area.  We begin, in this section, on the microscopic side (the
upper right corner of Fig.~\ref{fig:statecount}) by identifying the
most general supersymmetric bound states of branes that can be made
from 2-branes and 6-branes.    This can be done using the techniques
of~\cite{bdl,5charge}.

Working in lightcone frame,  $Q$ and $\tilde{Q}$, the two
supercharges of Type II string theory, are 16 component chiral $SO(8)$
spinors.  A Dp-brane imposes the projection~\cite{dnotes,bdl}
\footnote{We take $\Gamma_{11} \tilde{Q} = + \tilde{Q}$.}:
\begin{equation}
Q \pm \Omega_p(\gamma) \tilde{Q}= 0
\label{eq:susyproj}
\end{equation}
Here $\Omega$ is the volume form of the brane
$\Omega_p(\gamma) = {1\over (p+1)!}\epsilon_{i_0\cdots i_p} 
\gamma^{i_0} \cdots \gamma^{i_p} $,
and is normalized to be a projection. The $\pm$ signs in
Eq.~\ref{eq:susyproj} distinguish between branes and anti-branes or,
equivalently, between opposite orientations.\footnote{We will take the
torus to have unit moduli.  See~\cite{bdl} for extensions to a general
torus.}

In order to study simultaneous projections imposed by many 2-branes we
introduce complex coordinates $(z_1,z_2,z_3)$ on the 6-torus, related
to the real coordinates via $z_\mu=(y_{2\mu-1}+iy_{2\mu})/\sqrt{2}$,
$\mu=1,2,3$. The corresponding complexified Gamma matrices are
$\Gamma^{\mu}=(\gamma^{2\mu-1}+i\gamma^{2\mu})/2$ and their complex
conjugates $\bar{\Gamma}^{\bar
\mu}=(\gamma^{2\mu-1}-i\gamma^{2\mu})/2$.  Take $z_4$ and $\bar{z}_4$
to be complex coordinates for the remaining two dimensions in
lightcone frame and define $\Gamma^4$ and $\bar{\Gamma}^4$
accordingly.  These complexified matrices obey a Clifford algebra and
so define a Fock basis $|n_1,n_2,n_3\rangle \otimes | n_4\rangle$ on
which $\Gamma_\mu$ ($\bar{\Gamma}_{\bar\mu}$) and $\Gamma_4$
($\bar{\Gamma}_{\bar{4}}$) act as annihilation (creation) operators.
Specifically:
\begin{equation}
\bar{\Gamma}_{\bar\mu}\Gamma_{\mu}|n_1,n_2,n_3\rangle \otimes |n_4\rangle
=n_\mu|n_1,n_2,n_3\rangle \otimes |n_4 \rangle 
\label{eq:fock}
\end{equation}
where the $n_\mu$ and $n_4$ take values $0$ and $1$.   

\subsection{2-branes at Angles and 6-branes}
Using the above complex notation, a 2-brane wrapped on the $(y_1, y_2)$
cycle imposes the projection:
\begin{equation}
\gamma^0 Q =  \gamma^1 \gamma^2 \tilde{Q} = 
-i(2\bar{\Gamma}^{\bar 1}\Gamma^1 -1)\tilde{Q} 
\label{eq:susyrefconfig1}
\end{equation}
Consider a set of 2-branes rotated relative to this reference brane on
the 6-torus.  The ith brane is rotated by some $R_i \in SO(6)$ and
imposes the projection:
\begin{equation}
\gamma^0 Q =  (R_i\gamma)^1 (R_i\gamma)^2 \tilde{Q}
\end{equation}
where $R_i$ is in the fundamental representation of
$SO(6)$. \footnote{Of course, only a discrete subgroup of $SO(6)$ is
allowed so that winding numbers remain finite.}  We can also write
this as:
\begin{equation}
\gamma^0 Q = -i {\cal S}_{(R_i)} \: (2\Gamma^{\bar 1}\Gamma^1-1) \: 
{\cal S}^{\dagger}_{(R_i)} \:
\tilde{Q}
\label{eq:susyrotat1}
\end{equation}
where ${\cal S}_{(R_i)}$ is the spinor representation of the rotation.  The
Fock space elements $|n_1 n_2 n_3 \rangle \otimes |n_4\rangle$ which form a
basis for the spinors $\tilde{Q}$ are eigenstates of the $\Gamma$-matrix
projections in Eq.~\ref{eq:susyrefconfig1} and Eq.~\ref{eq:susyrotat1}:
$-i(2\bar{\Gamma}^{\bar j} \Gamma^j - 1) |n_1 n_2 n_3 \rangle \otimes |n_4
\rangle = i(1 - 2n_j) |n_1 n_2 n_3 \rangle \otimes |n_4\rangle$.  So there
are simultaneous solutions of Eq.~\ref{eq:susyrefconfig1} and all the
Eqs.~\ref{eq:susyrotat1} for each $i$, if there exist some $\tilde{Q}$
which are singlets under {\em all} the rotations: ${\cal
S}_{(R_i)}\tilde{Q} = {\cal S}^{\dagger}_{(R_i)}\tilde{Q} = \tilde{Q}$.

Given such a collection of $R_i \in SO(6)$ that leave some $\tilde{Q}$
invariant, the set of products of $R_i$ and their inverses form a subgroup
of $SO(6)$.  So the problem of finding supersymmetric relative rotations is
reduced to one of finding subgroups of $SO(6)$ that leave some $\tilde{Q}$
invariant.  Examining the explicit Fock basis in Eq.~\ref{eq:fock}, it is
clear that $|n_4\rangle$ is inert under $SO(6)$ rotations so that
$\tilde{Q}$ decomposes as ${\bf 16} \rightarrow {\bf 8}_0 + {\bf 8}_1$.
The ${\bf 8}$ indicates the 8 dimensional spinor representation of $SO(6)$
and the subscripts indicate the eigenvalue of $n_4$ in each representation.
The largest subgroup of $SO(6)$ under which spinors transform as singlets
is $SU(3)$, with the decomposition ${\bf 8} \rightarrow {\bf 1 + 3 +
\bar{3} + 1}$. So an arbitrary collection of 2-branes related by $SU(3)$
rotations is supersymmetric.  In fact, the branes can be related by $U(3)$
rotations because the $U(1)$ factor in $U(3) = SU(3) \times U(1)$ cancels
between ${\cal S}_{(R_i)}$ and ${\cal S}^\dagger_{(R_i)}$ in
Eq.~\ref{eq:susyrotat1}. Concisely, a collection of 2-branes may be wrapped
supersymmetrically on arbitrary $(1,1)$ cycles relative to some complex
structure.  Further global rotations of the entire configuration by
$SO(6)/U(3)$ rotations may be used to turn on 2-brane charges on arbitrary
2-cycles.  It is argued in~\cite{5charge} that the general supersymmetric
bound state of 2-branes on a 6-torus has this form.

    We can determine the amount of supersymmetry surviving the
presence of $U(3)$ rotated D2-branes by looking for $U(3)$ invariant
spinors $\tilde{Q}$.  Given the reference configuration
Eq.~\ref{eq:susyrefconfig1} and the Fock basis discussed above, it is
readily shown that the $U(3)$-invariant spinors are $\tilde{Q} = \{
|000\rangle \otimes |n_4\rangle, |111 \rangle \otimes |n_4\rangle\}$
where $n_4=\{0,1\}$.  These four solutions give the equivalent of
$N=1, d=4$ supersymmetry.  D6-branes can be added without breaking any
additional supersymmetry so long as we pick the orientation associated
with the minus sign in Eq.~\ref{eq:susyproj}.   The 6-brane imposes the
condition: 
\begin{equation}
\gamma^0 Q = -i(2\bar{\Gamma}^{\bar 1}\Gamma^1 - 1)
(2\bar{\Gamma}^{\bar 2}\Gamma^2 - 1)
(2\bar{\Gamma}^{\bar 3}\Gamma^3 - 1) 
\tilde {Q}
\label{eq:6proj}
\end{equation}
which is solved by the same spinors that survive the presence of the
2-branes. T-duality of the entire 6-torus converts the 6-branes into
0-branes.  Furthermore, 2-branes on some $(1,1)$ cycle turn into
4-branes on the dual $(2,2)$ cycle.  So we learn that 4-branes wrapped
on arbitrary $(2,2)$ cycles are supersymmetric since they are related
by relative $U(3)$ rotations, and 0-branes may be bound to them
without breaking supersymmetry.

\section{Classical Solutions for Branes at Angles}
\label{sec:classical}
Having identified supersymmetric microscopic configurations of
2-branes and 6-branes, we would like to find the corresponding
classical solutions.  These will turn out to be four dimensional
extremal black holes, giving us an opportunity to test the $E(7,7)$
invariant prediction for black hole entropy.

Choosing complex coordinates $z^j = (x^{2j-1} + ix^{2j})/\sqrt{2}$ on
the $T^6$, the K\"ahler form and volume 
of the asymptotic torus are
$k=i\sum_{J=1}^3 dz^{J} \wedge d\bar{z}^{\bar J}$
and ${\rm Vol}(T^6)=\int_{T^6}{\rm dVol} = \int_{T^6}
{\J\wedge\J\wedge\J / 3!}$.  The volume is set to $1$ by taking the
asymptotic moduli to be unity.\footnote{We follow the conventions
of~\cite{5charge} for the normalization of forms, wedge products and
Hodge dual.}  Following the previous section, we consider a
supersymmetric collection of 2-branes wrapped on $(1,1)$ cycles.
Thinking of each brane as being $U(3)$ rotated relative to the given
complex structure, the jth brane is characterized by the $(1,1)$
volume form
$\omega_j=i \, (R_{(j)})^1_J \, (R^*_{(j)})^1_K \, 
dz^{J} \wedge d\bar{z}^{\bar K}$.
Such wrapped branes produce a pressure on the geometry causing it to
deform between the position of the branes and the flat space at
infinity.  

Remarkably, the full solution can be understood in terms of
a single $(1,1)$ form $\omega$ characterizing the ensemble of branes: 
$\omega = \sum_j X_j \,\omega_j$
Here the $X_j$ are harmonic functions in the transverse space $X_j =
{P_j / r}$ and $P_j$ is the charge of the $j$th 2-brane.
\footnote{In general, we could separate the branes in the transverse
space by choosing $X_j = P_j/|\vec{r} - \vec{r}_j|$ but will not do so
here since we are interested in four-dimensional black holes.}  Since
the classical solutions will only depend on $\omega$, many different
microscopic configurations of branes will have the same macroscopic
solution - this is a reflection of the no-hair theorem.  It is also
useful to define the intersection numbers $C_{ij} =(1 /
\vol(T^6))\int_{T^6} \J\wedge \omega_i\wedge\omega_j$ and $C_{ijk}=(1
/ \vol(T^6))\int_{T^6} \omega_i\wedge \omega_j\wedge\omega_k$.
Then $C_{ijk}$ is proportional to the number of points at which a
T-dual collection of 4-branes intersect on the 6-torus.  This
connection will be used in Sec.~\ref{sec:counting} in computing the
degeneracy of the solutions constructed in this section.

A classical solution corresponding to a collection of 6-branes and
2-branes at angles on a 6-torus is completely described in terms of
the metric, the dilaton, the $RR$ 3-form gauge field and the $RR$ 7-form gauge
field. 
The solution in string metric is:
\begin{eqnarray}
ds^2&=&(F_2F_6)^{1\over 2} dx_\perp^2 + (F_2F_6)^{-{1 \over 2}} \left[-dt^2+
 (h_{\mu\bar\nu} \, dz^\mu\ d\bar{z}^{\bar\nu}  +
  h_{\bar\mu\nu} \, d\bar{z}^{\bar\mu}\ dz^{\nu}) \right]  \nonumber\\
A_{(3)}&=&{1\over F_2} \, dt\wedge K~~~~~~~~~~~;~~~~~~~~~~~
A_{(7)}=-{1\over F_6} \, dt\wedge {\rm dVol}
\label{eq:gaugeans}\\
e^{-2\Phi}&=&\sqrt{{F_6^3\over F_2}} \label{eq:phians}
\end{eqnarray}
where the 2-form $K$ is:
\begin{equation} 
K \equiv *{(k + \omega) \wedge (k + \omega) \over 2!}\label{eq:Kans}
\end{equation}
and is simply proportional to the internal K\"{a}hler metric in the
presence of 2-branes:
\begin{equation}
G = i g_{\mu\bar\nu} \, dz^{\mu} \wedge d\bar{z}^{\bar\nu} 
= {i \over \sqrt{F_2 F_6}} h_{\mu \bar\nu} \, dz^{\mu}
\wedge d\bar{z}^{\bar\nu} 
\equiv  { K \over \sqrt{F_2 F_6}} 
\end{equation}
The functions $F_2$ and $F_6$ have simple expressions:
\begin{eqnarray}
F_2&=&{\int_{T^6} (\J+\omega)^3\over 3!\vol(T^6)} 
= 1+\sum_i X_i +\sum_{i<j} X_iX_j \,C_{ij} +\sum_{i<j<k} X_iX_jX_k
\, C_{ijk} 
\nonumber 
\\
F_6 &=& 1 + {Q_6 \over r}
\label{eq:F6}
\end{eqnarray}
Here $dx_\perp^2 = dx_7^2 + dx_8^2 + dx_9^2$ refers to the noncompact
part of the space.

In~\cite{5charge} it is explicitly shown that this solution satifies the
spacetime equations of motion and is supersymmetric.\footnote{There is an
extensive literature on intersecting branes.  See~\cite{intersect}, for
example.  Classical branes at angles are discussed, among other articles,
in~\cite{angles}.  More references are provided in~\cite{5charge}.}  The
four dimensional Einstein metric is related to the string metric by
$e^{-2\Phi_4} = e^{-2\Phi} \sqrt{\det g_{int}} = \sqrt{ {F_6^3 / F_2}}
\sqrt{ {F_2 / F_6^3}} = 1$.  So the Einstein metric is: $ds_4^2 =
(F_2F_6)^{-1/2} (-dt^2) + (F_2 F_6)^{1/2} (dr^2 + r^2 d\Omega^2)$ which
describes a black hole with horizon at $r=0$.  The area of the sphere at
radius $r$ is $A = 4\pi r^2(F_2 F_6)^{1/2}$.  In the limit $r \rightarrow
0$ this gives the area:
\begin{equation}
A = 4\pi (Q_6\sum_{i<j<k} P_i P_j P_k C_{ijk})^{1/2}.
\label{eq:areageneral}
\end{equation}
In order to compare this with the prediction from duality it is
convenient to project the brane charges $P_j$ onto the different
2-cycles of the torus.  Defining a basis for $(1,1)$ forms
$\Omega^{a\bar b} = i\tf{a}{b}$, the charge on the $(a\bar{b})$ cycle
is given by $q_{a\bar b} = \sum_i P_i \alpha_{ia \bar b}$ where
$\omega_i=\sum_{a \bar b} \alpha_{i a \bar b}\Omega^{a \bar b}$.  In
terms of this charge matrix, the area in Eq.~\ref{eq:areageneral} can
be rewritten as:
\begin{equation}
A=4\pi\sqrt{Q_6\det q}
\label{eq:area2}
\end{equation}
Now consider the prediction from duality that $A = 4\pi\sqrt{J_4}$.
Comparing with Eq.~\ref{eq:pred} for $J_4$, all the $x_{ij}$ vanish
since there are no 4-branes or 0-branes.  Then, $y_{ij}$ for $i,j\leq
6$ represent 2-brane charges on different cycles while $y_{78}$ is the
6-brane charge.  Transforming the $J_4$ symbol into complex
coordinates as in this section, it becomes $J_4 = Q_6 \det{q}$
matching Eq.~\ref{eq:area2} for an arbitrary number of 2-branes at
angles in the presence of 6-branes.

\section{Counting the States of Our Black Holes}
\label{sec:counting}
The first step towards a microscopic understanding of the entropy of our
black holes is to understand how the physical charges $P_i$ and $Q_6$ are
related to the wrapping numbers of branes on different cycles.
Following~\cite{5charge}, the quantization condition for 6-branes is $Q_6 =
(l_s g/4\pi) N_6$ where $N_6$ is the number of times the 6-branes wrap the
entire torus and $l_s =2\pi \sqrt{\alpha'}$ is the string length.
Similarly, if $P_j$ and $\omega_j$ are the physical charge and volume form
characterizing the jth 2-brane, we have $P_j \omega_j = (l_s g/4\pi) M_j
v_j$ where $M_j$ is the wrapping number and $v_j$ is the element of {\it
integral} cohomology characterizing the cycle on which the brane is
wrapped.

We then define $N_{ijk} = {1 \over \vol(T^6)} \int_{T^6} v_i \wedge
v_j \wedge v_k$.  T-duality of the entire 6-torus turns 2-branes
wrapped on $v_i$ into 4-branes wrapped on $*v_i$.  Three 4-branes on a
6-torus generically intersect at a point and the $N_{ijk}$ are
integers counting the number of intersection points.  Using the
quantized charges we find that $ \sum_{i<j<k} P_i P_j P_k C_{ijk} =
({l_s g / 4\pi} )^3 \sum_{i<j<k} (M_i M_j M_k) N_{ijk} $ Finally, the
entropy of the black hole is given by $S/4G_N$ and the Newton coupling
is $G_N = g^2 \alpha'/8 = g^2 l_s^2/32\pi^2$.  Putting everything
together we find that:
\begin{equation}
S = 2\pi \sqrt{N_6 \sum_{i<j<k} (M_i M_j M_k) N_{ijk}}
\label{eq:entropy}
\end{equation}
We will now explain the microscopic origin of this entropy.

To count the states of our black holes it is easiest to dualize the entire
6-torus so that the 6-branes become 0-branes, and the 2-branes become
4-branes wrapped on the cycles $*v_j$.  Of course, the entropy in
Eq.~\ref{eq:entropy} remains the same, with $N_{ijk}$ counting the number
of points at which the ith, jth and kth 4-branes intersect.  In fact, it is
convenient to lift the system into 11 dimensions and view the 4-branes as
M-theory 5-branes intersecting along the 11th circle.  Imagine making the
torus small and the 11th circle big.  Then the mutual intersection of any
triple of 5-branes is a string that has $(0,4)$
supersymmetry~\cite{vf,juan1,5charge,edblack}.  Putting momentum along this
effective string would induce 0-brane charge from the 10 dimensional
perspective.  However, since we want to preserve supersymmetry we can only
introduce momentum in the left-moving direction.  From the D-brane point of
view, this means that only one of the two signs in Eq.~\ref{eq:susyproj}
is allowed, as observed for the 6-branes at the end of
Sec.~\ref{sec:angles}.

So the entropy of the black hole arises from the number of ways in which
$N_6$ units of left-moving momentum can be distributed amongst the
effective intersection strings.  It was argued heuristically
in~\cite{vf,juan1,behrndt2} that the left moving, non-supersymmetric sector
of the effective string has a central charge $c=6$.  As shown more
carefully in~\cite{edblack} there are indeed 6 bosons in the left moving
sector, corresponding to the position of the effective string on the $T^6$.
Actually, the authors of~\cite{edblack} count the states of black holes by
moving away from the singular limit of intersecting branes.  Instead they
consider a single M5-brane wrapped on $S^1$ times a complicated 4-cycle in
a Calabi-Yau 3-fold and count its fluctuations.  Their formulae do not
directly apply to tori which introduce some new features.  So we will not
use their techniques here and will remain in the intersecting brane limit
following~\cite{5charge}.

First consider a situation with $N_6 \gg \sum N_{ijk} M_i M_j M_k$.
Each of the $N_{int}=\sum N_{ijk} M_i M_j M_k$ effective strings has 6
bosons in its leftmoving sector.  So the problem is simply to
distribute $N_6$ units of momentum amongst $6 N_{int}$ bosons.  This
is identical to the problem of computing the density of states of a
string with central charge $c_{\rm eff} = 6 N_{int}$ and for large
$N_6$ this can be read off from~\cite{gsw}: $d(N_6)
=\exp{2\pi\sqrt{N_6 c_{\rm eff}/6}}$.  Taking the logarithm we find
the entropy:
\begin{equation}
S = 2\pi \sqrt{N_6 \sum_{i<j<k} (M_i M_j M_k) N_{ijk}}
\end{equation}
which exactly matches Eq.~\ref{eq:entropy}.  Here we assumed that at a
given intersection point of the ith, jth and kth branes there are $M_i
M_j M_k$ effective strings singly wound around the 11th circle.  We
can relax the condition $N_6 \gg \sum N_{ijk} M_i M_j M_k$ by
including the multiply wound strings arising from multiply wound
5-branes.

The easiest way to deal with this is to observe that a given triple
intersection of 5-branes with wrapping numbers $M_i$, $M_j$ and $M_k$
will be described by a $(0,4)$ supersymmetric sigma model on the
orbifold target:
\begin{equation}
{\cal M} = {(T^6)^{M_i M_j M_k} \over S(M_i M_j M_k)}
\label{eq:orbifold}
\end{equation}
We orbifold by the symmetric group $S(M_i M_j M_k)$ to account for symmetry
under exchange of the 5-branes.\footnote{One might have naively supposed
that the appropriate orbifold group would be $S(M_i)S(M_j)S(M_k)$.  This
would account for the exchange symmetry of each kind of 5-brane.  However,
the analysis of~\cite{sv1,dmvv} indicates that the appropriate group is
$S(M_1 M_2 M_3)$. Indeed, this is the orbifold that is consistent with
T-duality. }  We are interested in putting momentum on the leftmoving side
where there are 6 bosons characterizing the position on $T^6$.  In our
case, each triplet of 5-branes intersects in $N_{ijk}$ locations giving
rise to $N_{ijk}$ effective strings.  So from the M-theory perspective, our
solutions are described in the small torus limit by $N_{tot} = \sum_{i<j<k}
N_{ijk}$ effective strings, each propagating on an orbifold like
Eq.~\ref{eq:orbifold}.  The appropriate total effective conformal field
theories have central charges that are the sum of contributions from many
effective strings, and the resulting degeneracy exactly matches our
Eq.~\ref{eq:entropy} following~\cite{dmvv,dvv}.

\section{Discussion: 5 Parameter Generating Solution}  
\label{sec:discuss}
So far we have shown that duality prediction for the entropy of a black
hole is verified in detail for general supersymmetric systems of 2-branes
and 6-branes or 4-branes and 0-branes.  Classical solutions containing
arbitrary numbers of branes at angles have the appropriate horizon area and
it is possible to count the states microscopically.  However, as discussed
in Sec.~\ref{sec:gen}, the orbit under duality of the configurations
discussed so far does not account for the full 56 charge spectrum of four
dimensional black holes.  One way to see this is to observe that the
entropy in Eq.~\ref{eq:area2} is characterized by four invariant parameters
- $Q_6$ and the 3 real eigenvalues of the charge matrix $q$.  Equivalently,
in the notation of Sec.~\ref{sec:n8}, we have studied configurations in
which the charge matrix $z_{ij}$ is either pure real or pure
imaginary.

The additional invariant phase in the charge matrix that is absent in our
solutions is associated with the presence of electric-magnetic dual pairs
at the same time.  It is therefore natural to expect that the desired
generating solution can be constructed solely out of self-dual
3-branes.\footnote{In writing this section I have benefitted from Finn
Larsen's unpublished notes concerning the $E_7$ quartic invariant and black
holes.}  Due to lack of space, the corresponding classical solutions and
the counting of states will be discussed elsewhere.  Here we show how
3-branes at angles can be a convenient generating configuration and derive
the duality prediction for the entropy.

Let $z^i = (x^{2i-1} + i x^{2i})/\sqrt{2}$ be complex coordinates for
a 6-torus. Consider any collection of 3-branes wrapped on cycles ${\rm
Re}(\tilde{z}^1 ,\tilde{z}^2 ,\tilde{z}^3)$, with $\tilde{z}^i = R^i_j
z^j$ where $R \in SU(3)$.  Such a collection of 3-branes is
supersymmetric~\cite{bdl}.  We consider diagonal $SU(3)$ rotations $R=
{\rm Diag}(e^{-i\theta_1},e^{-i\theta_2},e^{-i\theta_3})$ that rotate
the branes separately on the $(12)$, $(34)$ and $(56)$ tori.  The
generating configuration has the following branes at angles:
\begin{table}[h]
\caption{Generating Configuration \label{anglesfig}}
\begin{center}
\begin{tabular}{cccc}
\hline
Wrapping Numbers     & $\theta_1$       & $\theta_2$      & $\theta_3$       \\
\hline
$N_1$ &  $0$             & $0$             &  $0$             \\
$N_2$ &  $0$             & ${\pi / 2}$ & ${-\pi / 2}$ \\
$N_3$ & ${-\pi / 2}$ & $0$             & ${\pi / 2}$  \\
$N_4$ & $\theta$         & $-\theta$       & $0$              \\
\hline
\end{tabular}
\end{center}
\end{table}

Here $\theta_i$ is the angle on the $(x^{2i-1},x^{2i})$ torus and
$N_j$ are the wrapping numbers on the rotated 3-cycles.  In order that
all wrapping numbers are finite, $\cot\theta = p/q$ is rational.  If
$N(stu)$ is the induced 3-brane charge on the $(stu)$ cycle we have:
$N(135) = N_1 + N_4 p^2$, $N(146) = -N_2$, $N(236) = -N_3$, $N(245) =
-N_4 q^2$, $N(145) = -p q N_4 $, $N(235) = p q N_4 $.

To show that this configuration displays the five parameters required
in a generating solution, we T-dualize on the 1, 3 and 5 cycles.  The
first three sets of branes turn into 0-branes and 4-branes, and the
last set turns into 4-branes with worldvolume fluxes turned on.  The
worldvolume fluxes induce additional 0-brane and 2-brane charges via
the Chern-Simons couplings on the 4-brane~\cite{ibrane}.  The
net charges can be read off from the $N(stu)$ 3-brane charges.  We
have $Q_0 = N_1 + N_4 p^2$ 0-branes, and $-N_4 q^2$ 4-branes on the
(1234) cycle.  There are $-N_3$ and $-N_2$ 4-branes on the (1256) and
(3456) cycles with $-N_4 pq$ and $N_4 pq$ 2-branes on the dual (34)
and (12) cycles.  Assembling these charges into the $J_4$ invariant in
Eq.~\ref{eq:pred} we find:
\begin{equation}
J_4 = Q_0 Q_2 Q_3 Q_4 - {Q_4^2 \cot^2\theta \over 4} (Q_2 + Q_3)^2 
\end{equation}
where $Q_2 = N_2$, $Q_3 = N_3$. $Q_4 = N_4 q^2$ and $\cot\theta = p/q$.
The second term in this equation arises from the mixing of the real and
imaginary parts of the charges $z_{ij}$.  This was not probed by the
configurations studied in earlier sections.

To verify that this is a five parameter generating configuration we
transform to the $SU(8)$ representation of the charge matrix following
Eq.~\ref{eq:SU8form}.  It is helpful to choose variables $P_L = Q_3 -
Q_2$, $P_R = -(Q_3 + Q_2)$, $Q_L = Q_0 - Q_4$, $Q_R = Q_0 + Q_4$ and
$\beta = Q_4 \cot\theta$.  Then the $SU(8)$ form of the charge matrix
is:
\begin{eqnarray}
Z_{12} & = &{1\over 4}(Q_L -P_L -2i\beta)~~~~~~~~;
~~~~~~~~Z_{56}={1\over 4} (Q_R + P_R) \\
Z_{34} & = &{1\over 4}(Q_L +P_L +2i\beta)~~~~~~~~;
~~~~~~~~Z_{78}={1\over 4} (Q_R - P_R) 
\end{eqnarray}
The resulting $J_4$ invariant is:
\begin{equation}
          J_4 =  \left[ \left({Q_R^2 - Q_L^2 \over 4}\right)
                   \left({P_R^2 - P_L^2 \over 4}\right)
                 - \left({\beta^2 P_R^2 \over 4}\right) \right]
\end{equation}
Written this way, the charge matrix and $J_4$ invariant are precisely
the same as those of the NS-NS, five parameter generating
solution discussed in~\cite{ct2,hull96}.  This shows that the
classical solution for the configuration of 3-branes at angles in
Table~\ref{anglesfig} is also a generating solution.

A definitive way of counting the states of this generating configuration is
to extend the analysis of~\cite{edblack} to tori.  The 4-branes then become
M5-branes wrapped on a circle and the 2-brane charges become flux on the
5-brane.  We hope to report on this elsewhere.  Counting the states of our
generating configuration accounts, after duality, for the entropy of the
entire 56 parameter space of extremal black holes in $N=8$ supergravity.

\acknowledgements 

I am grateful to R. Gopakumar, F. Larsen and A. Strominger for useful
conversations and comments.  Much of this work was done in
collaboration with F. Larsen and R. Leigh.  I thank the organizers of
the 1997 Cargese School for the opportunity to present it.  My
references are not intended to be comprehensive and I apologize for
omissions.  I am supported by the Harvard Society of Fellows and by
the NSF under grant NSF-Phy-91-18167.



\end{document}